\documentclass[a4paper]{jpconf}
\usepackage{graphicx}
\usepackage{amsmath,amssymb,amsfonts}
\usepackage{subfigure}

\newcommand{\mom}[1]{\left\langle{#1}\right\rangle}
\newcommand{\cum}[1]{\left\langle\!\!\left\langle{#1}\right\rangle\!\!\right\rangle}

\begin{document}
\title{Statistics of the quantized microwave electromagnetic field in mesoscopic elements at low temperature}

\author{St{\'e}phane Virally, Jean Olivier Simoneau, Christian Lupien and Bertrand Reulet}

\address{Universit{\'e} de Sherbrooke, Physics Department,\\2500, Boulevard de l'Universit{\'e}, Sherbrooke, QC J1K 2R1, Canada}

\ead{Stephane.Virally@USherbrooke.ca}

\begin{abstract}
The quantum behavior of the electromagnetic field in mesoscopic elements is intimately linked to the quantization of the charge. In order to probe nonclassical aspects of the field in those elements, it is essential that thermal noise be reduced to the quantum level, i.e. to scales where $kT\lesssim h\nu$. This is easily achieved in dilution refrigerators for frequencies of a few GHz, i.e. in the microwave domain. Several recent experiments have highlighted the link between dsicrete charge transport and discrete photon emission in simple mesoscopic elements such as a tunnel junction. Photocount statistics are inferred from the measurement of continuous variables such as the quadratures of the field. 
\end{abstract}

\section{Introduction}

Although charge currents in conductors are transported by fermions, they are always linked to bosonic excitations of the electromagnetic field. At the mesoscopic scale, currents exhibit properties that reflect the quantization of the electric charge (such as electron shot noise~\cite{Blanter2000}), and it is reasonable to expect that the electromagnetic field also behaves in a quantum manner, i.e. that photon shot noise should be observed~\cite{Beenakker2004, Girvin2009, Lebedev2010}. This is indeed the case and it can be observed in conditions where thermal noise is reduced to or below the level of the vacuum fluctuations of the field. Such conditions are met for microwave signals of a few GHz emitted by mesoscopic elements cooled down to a few milliKelvin, typically in dilution refrigerators~\cite{Grimsmo2016}.

Photon counters are difficult to come by in the microwave domain~\cite{Romero2009, Govenius2016}, as there are no natural systems with the right energy gaps. On the other hand, continuous observables such as the voltage can be measured with precision~\cite{Virally2016}. This contrasts with the case of the electromagnetic field in the visible and near infrared spectra, where the field itself is difficult to measure directly. Consequently, mesoscopic systems at low tempreature constitute a very interesting field of study for continuous variable quantum optics. This paper details some of the techniques required for the study of statistical properties of the quantized electromagnetic field in the microwave domain. It summarizes a recent series of experiments on one of the simplest mesoscopic systems, a tunnel junction cooled down to $\sim$20~mK and biased with both dc and ac signals~\cite{Gasse2013, Gasse2013a, Forgues2013, Forgues2014, Forgues2015, Simoneau2016}. It also shows that questions surrounding wideband signals, that are hard to answer when studying quantum optics at room temperature with visible or infrared signals, can be studied in details in mesoscopic systems.

\section{Narrowband signals}
The main observable in electric circuits is the voltage
\begin{equation}
\hat{v}(t)=i\sqrt{\frac{2}{Z}}\int_{0}^{+\infty}d\nu\;\sqrt{h\nu}\left[\hat{a}^\dagger(\nu)\,e^{i2\pi\nu t}-\hat{a}(\nu)\,e^{-i2\pi\nu t}\right],
\end{equation}
where $Z$ is the characteristic impedance of the transmission line in which the electromagnetic field propagates, and $\hat{a}^{(\dagger)}(\nu)$ are the usual bosonic annihilation (creation) operators.
When the detection bandwidth is narrow, the voltage observable, viewed in the rotating frame (i.e. with the use of a mixer), reduces to a single quadrature of the field,
\begin{equation}
\hat{x}_\theta=\frac{\hat{a}^\dagger\,e^{i\theta}+\hat{a}\,e^{-i\theta}}{\sqrt{2}},
\end{equation}
with $\theta$ a phase angle that can be selected with a phase shifter on the local oscillator signal. It can also be continuously changed by using a local oscillator slightly detuned from the signal. In other cases, the phase varies randomly during detection. In any case where the phase $\theta$ can be averaged, expectations of all odd centered moments of the measured signal are zero, while expectations of the $2k^\textrm{th}$ centered moments are
\begin{equation}
\label{X2k}
\mom{x_\theta^{2k}}_\theta=\left(\frac{1}{2}\right)^k\mom{\sum_\textrm{c.s.}\hat{a}^k\,\hat{a}^{\dagger k}},
\end{equation}
with ``c.s." standing for completely symmetric (i.e., the sum is taken on all possible permutations of the non-commuting operators). Here, $\mom{\bullet}$ represents averaging over the quantum ensemble, while $\mom{\bullet}_\theta$ represents averaging over both the quantum ensemble and the phase $\theta$. To simplify notations, we drop the $\theta$ indices everywhere below.

The completely symmetric sum of Eq.~\ref{X2k} can be explicitely evaluated and yields~\cite{Cahill1969a}
\begin{equation}
\sum_\textrm{c.s.}\hat{a}^k\,\hat{a}^{\dagger k}=\sum_{i=0}^k\left(\frac{1}{2}\right)^{k-i}\frac{(2k)!}{(i!)^2(k-i)!}\;\prod_{j=0}^{i-1}\left(\hat{n}-j\right),
\label{sumG}
\end{equation}
with $\hat{n}=\hat{a}^\dagger\,\hat{a}$ the usual number operator. Hence, the even moments $\mom{x^{2k}}$ of the continuous distribution are explicitely linked to the moments $\mom{n^\ell}$ of the discrete photocount distribution. As a measurement of $\mom{x_\theta^{k}}$ for all $k$ and all $\theta$ enables a full reconstruction (tomography) of the quantum state, the less stringent measurement of $\mom{x^{2k}}$ for all $k$ leads to a full reconstruction of the photocount distribution. This is expected, as we simply lose the phase information.

It is usually better to compute the cumulants of a distribution, rather than its moments, as cumulants are additive for independent distributions. Hence, noise cumulants can be readily subtracted from signal cumulants. The first moments of the photocount distribution are obtained from the first cumulants of the continuous voltage distribution as
\begin{align}
&\mom{n}=\cum{x^2}-\frac{1}{2};\label{n}\\
&\mom{\delta n^2}=\frac{2}{3}\cum{x^4}+\cum{x^2}^2-\frac{1}{4};\label{n2}\\
&\mom{\delta n^3}=\frac{2}{5}\,\cum{x^6}+4\cum{x^4}\cum{x^2}+2\cum{x^2}^3-\frac{1}{2}\,\cum{x^2}\label{n3},
\end{align}
where $\cum{\bullet}$ are the cumulants, as opposed to the central moments ($\mom{\bullet}$) of the distribution.

\begin{figure}
\centering
\subfigure[Experimental determination of the first moments of the photocount distribution for a coherent state.]{\includegraphics[width=0.45\textwidth]{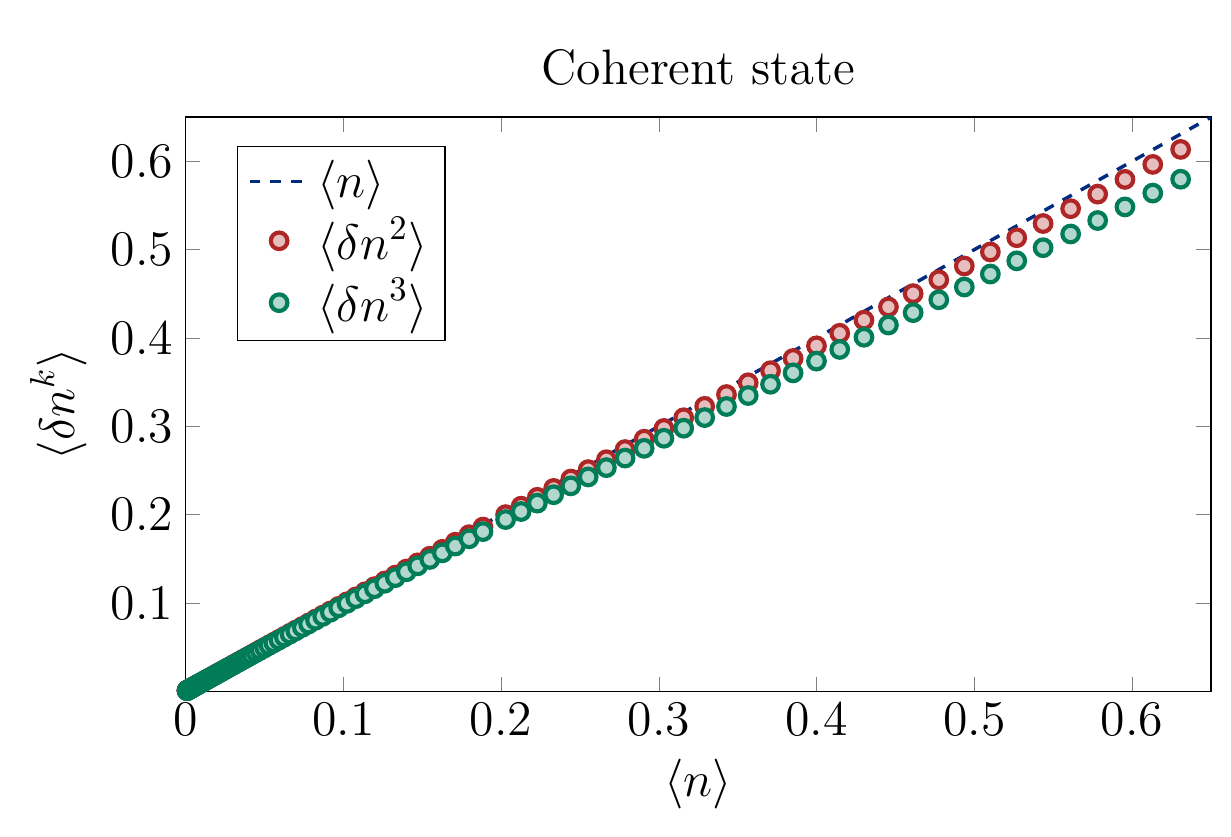}}
\hspace{0.05\textwidth}
\subfigure[Classical and nonclassical domains in the $\left(\mom{n},\mom{\delta n^2}\right)$ plane.]{\includegraphics[width=0.45\textwidth]{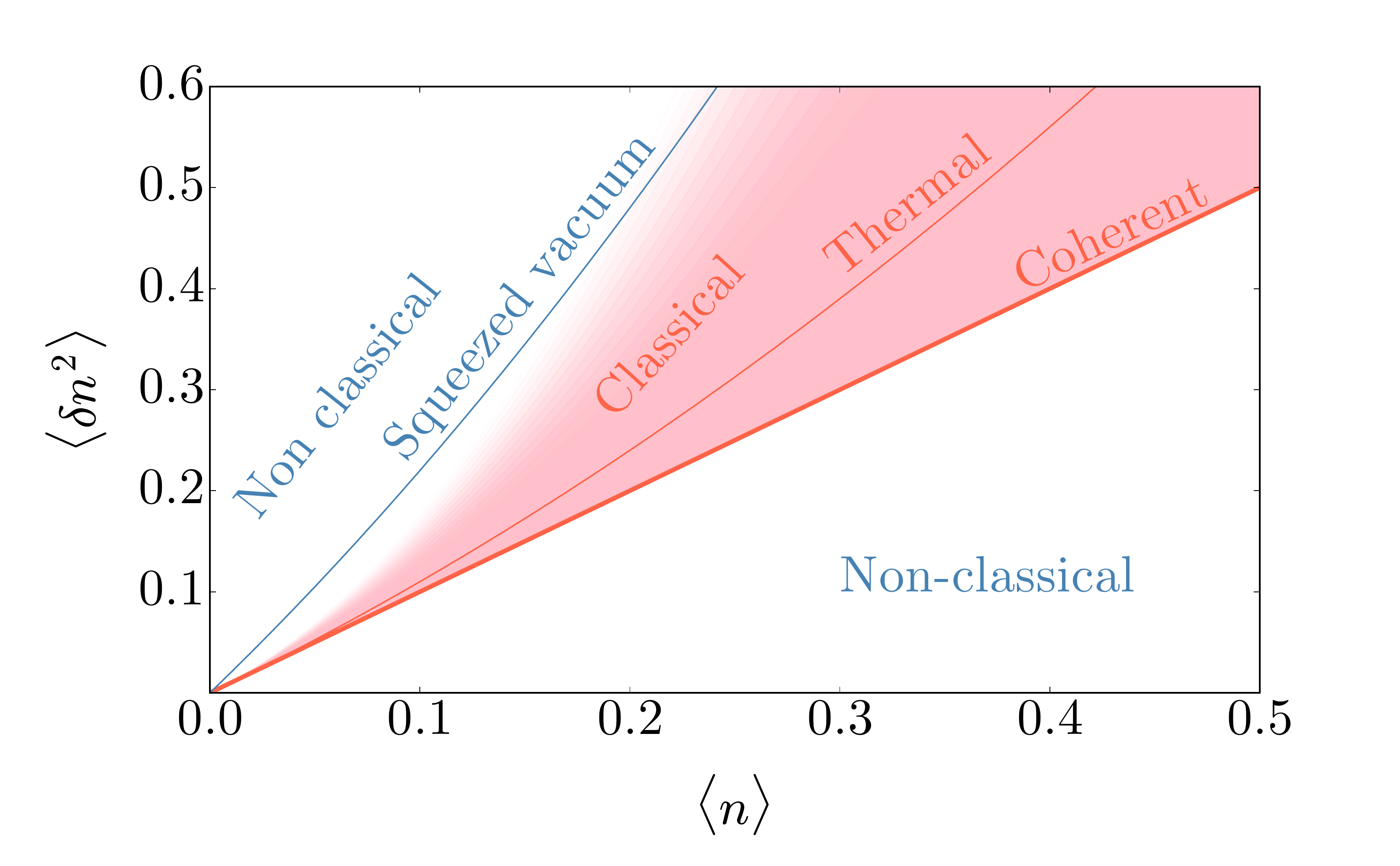}}
\caption{$\mom{\delta n^k}$ vs. $\mom{n}$.}
\label{C/NC}
\end{figure}

As a proof-of-principle experiment, we measured the first three moments of the photocount distribution of a narrow-band coherent signal (i.e. a sine wave) in the limit $\mom{n}<1$~\cite{Virally2016}. In order to do so, we had to place an ``emitter" (the last attenuator of a long transmission line) in a cryogenic environment to ensure $h\nu\gg k_BT$, with $\nu$ the frequency of the excitation ($\sim$ 6GHz) and $T$ the temperature ($\sim$ 10mK). The results are shown on Fig.~\ref{C/NC}a. For such a state, we expect $\mom{\delta n^2}=\mom{\delta n^3}=\mom{n}$.

Coherent states are only one variety of states that exhibit classical properties for the photocount distribution. Among those properties are
\begin{align}
&\mom{\delta n^2}\ge\mom{n};\\
&\mom{\delta n^2}\rightarrow\mom{n}\textrm{ as }\mom{n}\rightarrow 0;\\
&\mom{\delta n^3}\ge\mom{n}+3\left(\mom{\delta n^2}-\mom{n}\right)\left(1-\mom{n}\right).
\end{align}
In particular, it is possible to show domains of nonclassicality on the $\left(\mom{n},\mom{\delta n^2}\right)$ plane (see Fig.~\ref{C/NC}b).

\begin{figure}
\centering
\subfigure[Tomography of the squeezed vacuum]{\includegraphics[width=0.45\textwidth]{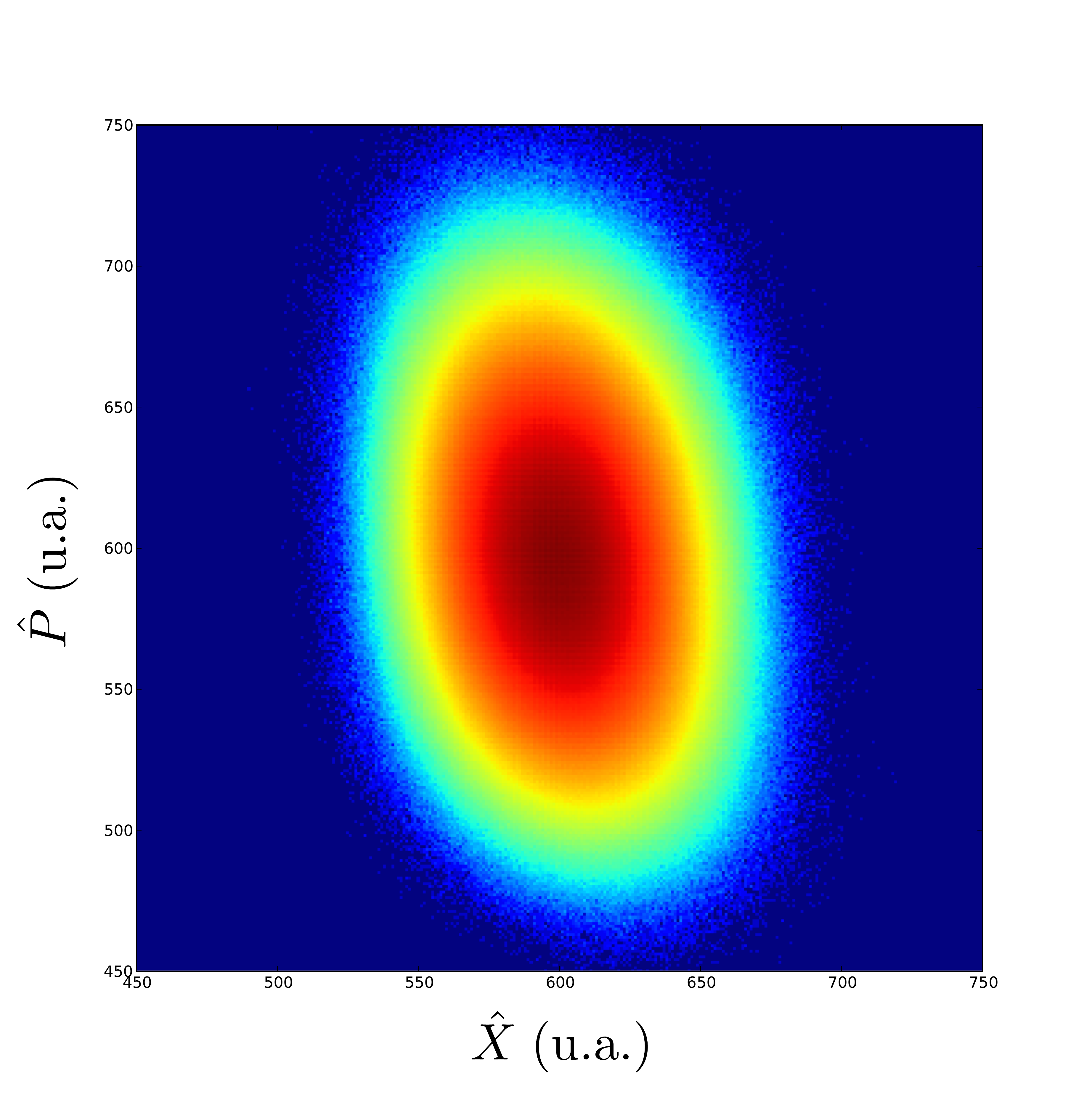}}
\hspace{0.05\textwidth}
\subfigure[Noise in both quadratures.]{\includegraphics[width=0.45\textwidth]{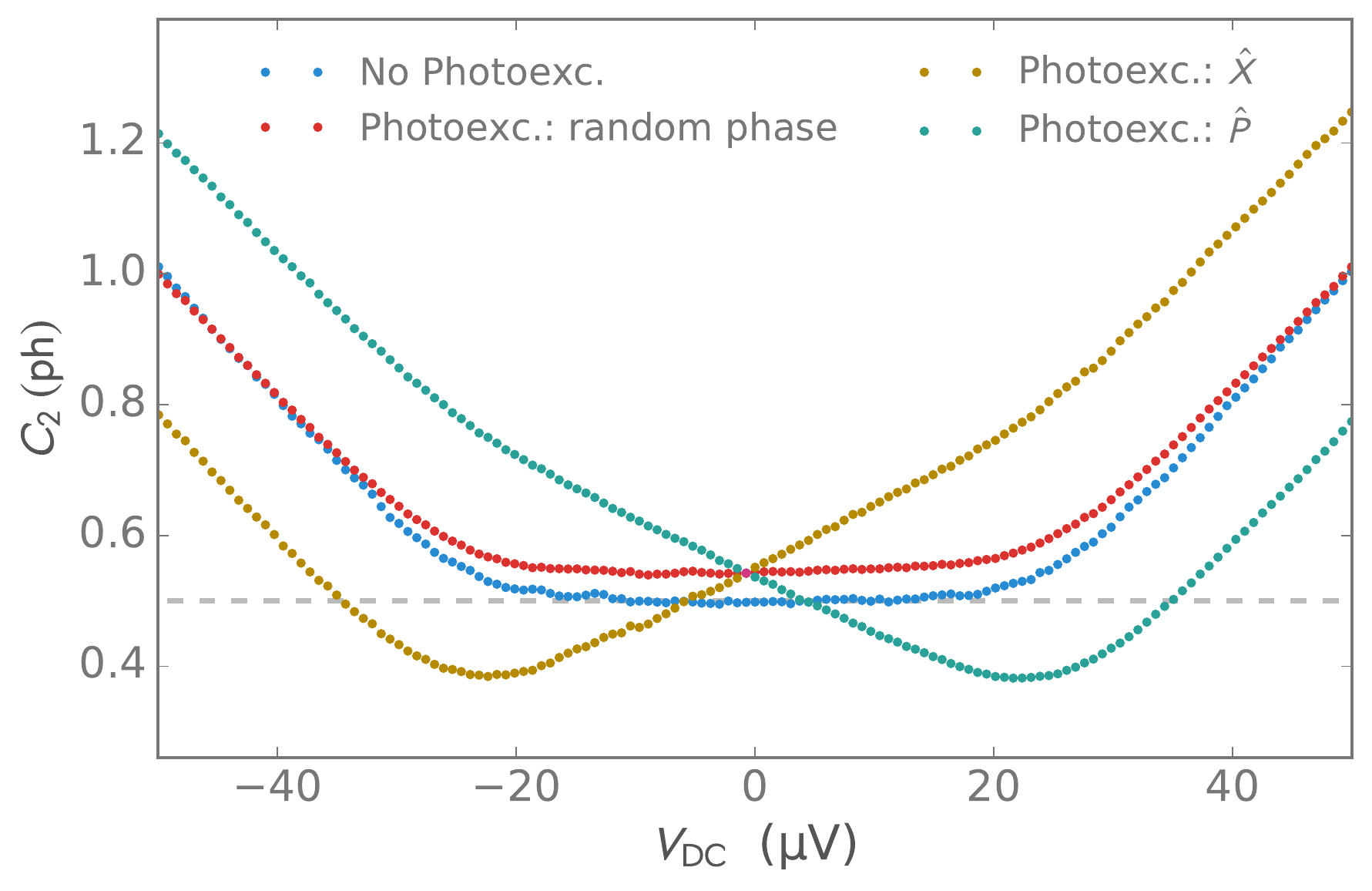}}
\caption{Squeezed vacuum state produced by an ac+dc bias of a tunnel junction.}
\label{JO}
\end{figure}

The noise emitted by a tunnel junction biased in dc and ac exhibits some of the nonclassical characteristics defined above. In the narrow band case at half the frequency of the ac excitation, this noise exhibits a Fano factor $\mathcal{F}=\mom{\delta n^2}/\mom{n}$ greater than unity, even in the limit of vanishing $\mom{n}$ (at least at $T=0$), in contradiction with the second classical condition above. This was demonstrated in a recent series of experiments showing that the electromagnetic noise is the result of a squeezing of the vacuum modes of the electromagnetic field~\cite{Gasse2013, Gasse2013a, Forgues2013, Forgues2014, Forgues2015}. In particular, a Fano factor above unity was observed in Ref.~\cite{Simoneau2016}. Squeezing is produced when pairs of photons are generated simultaneously, which is the root of the greater than unity Fano factor (it would be 2 for a pure source of photon pairs, down to $\mom{n}=0$ at $T=0$). The signature of squeezing is an asymetry between the noise in two orthogonal quadratures ($\hat{x}_\theta\equiv\hat{x}$ and $\hat{x}_{\theta+\pi/2}\equiv\hat{p}$). Compared to the symetric vacuum noise, squeezed signals sub-vacuum noise in one quadrature, at the expense of added noise to the orthogonal quadrature. This is illustrated in Fig.~\ref{JO}.

\section{Wideband signals}
The voltage is directly proportional to the electric field. In regular quantum optics, the field itself is difficult to probe as its oscillations are much faster than the response time of the detectors. In contrast, mesoscopic systems at low temperature offer great opportunities to directly probe field oscillations in real time, as the GHz signals required for these experiments are directly accessible through fast sampling oscilloscopes. Interestingly, one can even probe a wideband signal with a setup allowing detection from dc to a few GHz. This is not possible in visible or near infrared quantum optics, where detection bands are always narrow compared to the value of the central frequency of the signal. Of course, probing of the quantum behavior of such a signal requires that $k_BT\ll h\Delta\nu$, where $\Delta\nu$ is the bandwidth of the signal.

For wideband signals, it is not immediately obvious what a photon is. A single excitation of the electromagnetic field is in a superposition of many wavelengths, and the question of how many photons are present in a single pulse is not trivial. However, one can imagine an experiment that determines such a number, and there must be a theory predicting the results. The thought experiment is as follows: imagine a gapless material (e.g. graphene) absorbing photons from a short, wideband pulse. All frequencies in the pulse can lead to an electron being ejected from the valence band and injected into the conduction band, before being amplified by a cascading process. Hence, it is theoretically possible to ``count'' the number of photons in such a pulse. If one were to carry such an experiment, all information on the energy of each photon would be lost, as dictated by Heisenberg's time/energy uncertainty principle. On the other end, it is possible to simply measure the energy content of the pulse, with a bolometer for instance. In this case, all timing information would be erased.

\begin{figure}
\centering
\includegraphics[width=0.95\textwidth]{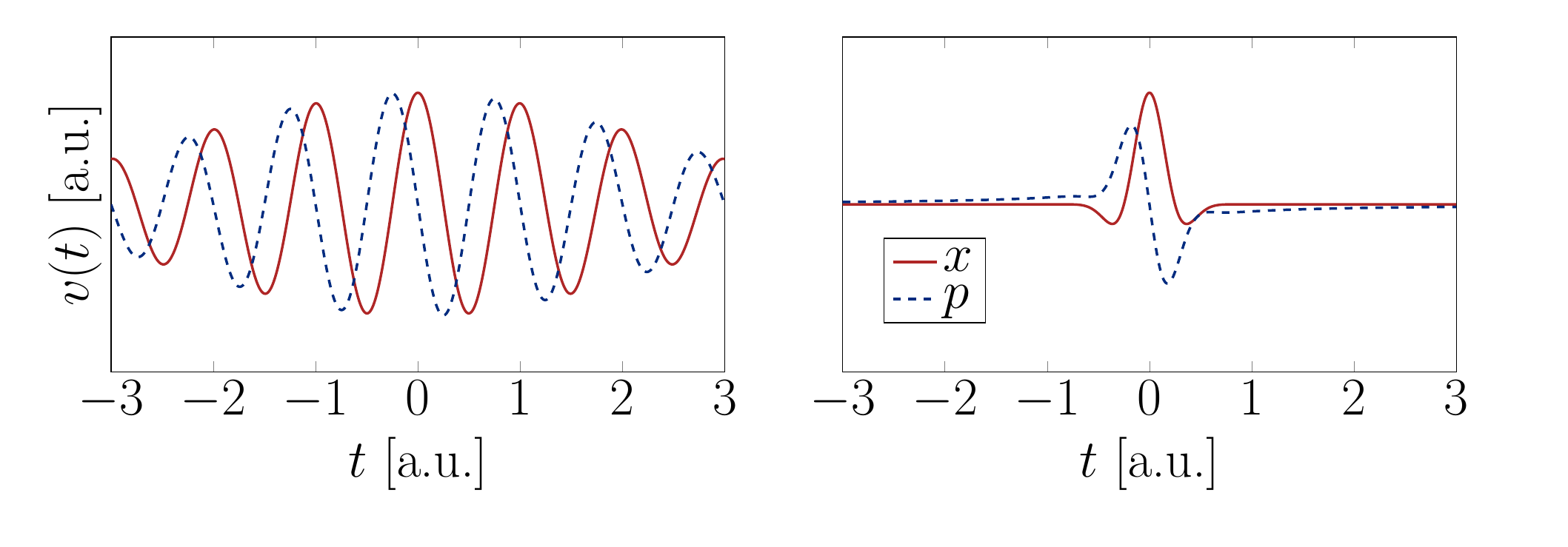}
\caption{Quadratures of narrow band (left) and wideband (right) signals.}
\label{XP}
\end{figure}

As in the narrow band case, it is in fact possible to obtain information on the continuous energy distribution and on the discrete photocount distribution by simply measuring the voltage on a sufficiently fast oscilloscope. It is found that the energy observable is simply
\begin{equation}
\frac{1}{Z}\int_{-\infty}^{+\infty}dt\;\hat{v}^2(t)=\int_0^{+\infty}d\nu\;h\nu\;\hat{a}^\dagger(\nu)\hat{a}(\nu),
\end{equation}
while the observable for the number of photons is
\begin{equation}
\frac{2}{h\,Z}\int_{-\infty}^{+\infty}dt\;\hat{w}^2(t)=\int_0^{+\infty}d\nu\;\hat{a}^\dagger(\nu)\hat{a}(\nu),
\end{equation}
where we have defined the causal transform
\begin{equation}
\hat{w}(t)=\int_0^{+\infty}d\tau\;\frac{\hat{v}(t-\tau)}{\sqrt{\tau}}.
\end{equation}
It turns out that the number of photons can always be written as a sum of the squares of two quadratures
\begin{equation}
\frac{1}{h\,Z}\int_{-\infty}^{+\infty}dt\left[\hat{x}^2(t)+\hat{p}^2(t)\right],
\end{equation}
and that these quadratures are Hilbert transforms of one another ($\hat{w}$ is one of these quadratures). Fig.~\ref{XP} illustrates the quadratures of signals in the narrow band and wideband limits. In the narrow band limit, we find, as expected, that the quadratures are essentially the sine and cosine at the central frequency.

\section{Conclusion}
Mesoscopic systems at low temperature provide interesting platforms for quantum optics. Time-resolved voltage measurements can fully characterize quantum states of the electromagnetic field, on the condition that the emitters be at low enough temperature that the condition $h\nu\gg k_BT$ (narrow band), or $h\Delta\nu\gg k_BT$ (wideband) be respected. Some mesoscopic conductors have already been shown to exhibit purely quantum behaviors. Such is the case for a tunnel junction appropriately biased in ac and dc. Interestingly, quantum optics in the microwave regime also offers interesting perspective for time domain quantum optics, as the fluctuations of the field can be fully resolved in time, contrary to visible or near-infrared optics. We have shown that the tools already exist for this exploration, and that exciting new physics can be probed in the near future in those systems.

\section*{Acknowledgements}
This work was supported by the Canada Excellence Research Chairs program, Canada Foundation for Innovation, Canada NSERC, Qu\'{e}bec MEIE, Qu\'{e}bec FRQNT via INTRIQ, and Universit\'{e} de Sherbrooke via EPIQ. 

\section*{References}
\bibliographystyle{iopart-num}
\bibliography{Bib}

\end{document}